# CONTEXT MANAGEMENT STRATEGIES IN WIRELESS NETWORK

Giadom, Vigale Leelanubari
M.Sc Student, Department of Computer Science
University of Port Harcourt
Rivers State, Nigeria

Dr. E. E. Williams
Department of Computer Science
University of Calabar
Cross River, Nigeria

*Abstract*—Context management strategies in wireless technology are dependent upon the collection of accurate information from the individual nodes. This information (called context information) can be exploited by administrators or automated systems in order to decide on specific management concerns. While traditional approaches for fixed networks are more or less centralized, more complex management strategies have been proposed for wireless networks, such as hierarchical, fully distributed and hybrid ones. The reason for the introduction of new strategies is based on the dynamic and unpredictable nature of wireless networks and their (usually) limited resources, which do not support centralized management solutions. In this work, efforts is being made to uncovered some specific strategies that can be used to managed context information that reaches the centre of decision making, the work is concluded with a detail comparison of the strategies to enable context application developers make right choice of strategy to be employed in a specific situation.

## I. INTRODUCTION

Researches in context-aware computing have given rise to several application models, theories, principles, prototype, frameworks and middleware systems for describing context but the development of context-aware programs and applications is still a difficult task [13]. This stems from the fact that there is a limited or no knowledge on the set of strategies to be employed for optimal performance and to enhance the efficiency of a context application. Also current middleware applications require a global knowledge of the context-aware infrastructures in order to establish context based interactions [13]. This project seeks to explore Context Management Strategies in wireless Technology.

A handful of research has been done on context management, although most of these works center on context management in mobile and distributed network but little or no work has being done on the strategies to be employed in managing the context information that comes into a context manager. Christine (2004) revealed that a context management system is a distributed system that enables applications to obtain context information. These information forms a key component of any pervasive computing. HE FURTHER CATEGORIZES CONTEXT MANAGEMENT INTO:

A. MOBILE ENVIRONMENT

B. HOME/OFFICE ENVIRONMENT

C. ADHOC ENVIRONMENT

Ilka (2001) stated that Context management at the access Reuters is considered to support establishment, updating and deleting of the mobile nodes context information for a particular service, aimed for instance at multicast listening, Quality of Service (QoS), reservation, accounting and security.

In all these works, Context Management Strategies in Wireless Technology was not really captured. The dearth of materials, research works and active research groups on Context Management Strategies in Wireless Technology have necessitated this project. This project captures a practicable area of computing called networking, specifically wireless technology. A research on Context Management Strategies in wireless technology is necessary as the world is gradually moving away from wired network to a more reliable, cable less network. This cable less network is convenient, flexible, fast, dynamic, secured and has a wider coverage.

## II. CONTEXT ANALYSIS

### A. CONTEXT IN NETWORK ENVIRONMENT

Schilit et al, 1994, the pioneers of context-aware computing, regard context to be location, identities of nearby people and object, and changes to those objects. They consider where you are, whom you are with, and what resources are nearby to be the important aspects of context. Abowd et al.'s more recent classification of context (Abowd et al, 1997) expands the Schilit et al, 1997 definition. They define context as: "…any information that can be used to characterize the situation of an entity. An entity is a person, place or object that is considered relevant to the interaction between a user and an application, including the user and application themselves." This means that any information that depicts the situation of a user can be entitled context. The temperature, the presence of another person, the nearby devices, the devices a user has at hand and the orientation of the user are examples.

### B. CONTEXT MANAGENENT

Context management encompasses all the processes involved in the collection, exchange and processing of context information. This is necessary because:
- The context information can be distributed in both network and mobile terminal. Therefore, it is difficult to collect and need to be exchanged.
- Wireless link is a bottle neck for context exchange. Hence the information exchange between the mobile node and networks should be minimized.
- The context information can be either dynamic or static. The dynamic information needs to be updated frequently.

Moreover, the amount of context information can be enormous. For a specific service, some context information is relevant, some is not relevant. Therefore, we need a data structure to compile the context information effectively.

### C. CONTEXT INFORMATION

Context information is data that describes the state of a certain entity at a specific moment. Context information makes applications and networks aware of their situation and improves certain functions in a network or applications, but is typically not critical in the sense that functionality cannot be provided at all if the context information is not available. However, context information is expected to play a vital role in supporting autonomic decision making and is regarded as a necessity for self-organization [3].

Basically, there are two classifications of context information which are Static Context Information and Dynamic Context Information.

### D. CONTEXT MANAGEMENT SYSTEMS

A context management system is either a centralized system that enables applications to obtain context information about (mobile) users and forms a key component of any pervasive computing environment (Christine, 2004). Context management systems are however very environment-specific (e.g., specific for home environments) and therefore do not interoperate very well. This limits the operation of context-aware applications because they cannot get context information on users that reside in an environment served by a context management system that is of a different type than the one used by the application. This is particularly important for mobile users, whose context information is typically available through different types of context management systems as they move across different environments. A context management system is a computational element responsible for binding context providers, which produce context information, and context consumers, typically represented by context-aware applications. The main task of a context management system is to match consumer's interests with probed context information. The complexity of context management in a distributed scenario is defined by the wideness of an interest, i.e. the number of context management systems that should be involved in an interest matching. If a distributed scenario is also open, heterogeneous and dynamic then the wideness of an interest is variable, as a result of characteristics such as dynamic introduction of new sensors and evolution of context models. The support of context interest of variable wideness imposes challenging requirements for context management systems [16].

### E. CONTEXT AWARE PROGRAMS

These are set of instructions, commands and applications that can adapt and perform tasks on ambient conditions in the physical or the virtual world. In context-aware applications, adaptations are triggered by changes of certain context information. For example, smart applications designed to support meetings may automatically transfer a presentation to a projector as soon as the presenter enters the meeting room. In this case, both the location of the presenter and his/her role in the meeting room are basic pieces of context information used to trigger the transfer of the presentation. Basically, the development of a context-aware application, as in this example, involves the description of the actions to be triggered according to a set of contextual conditions. A same piece of context information may be used for different purposes. The location of the presenter, for example, may also be used by another application to disseminate his availability status for an instant communicator. Moreover, this context information may be provided by different sensors, such as a proximity sensor to identify if the user is inside the classroom and using a microphone connected to voice recognition software to identify specific users in the classroom [16].

### III. CONTEXT IN A WIRELESS NETWORK

Various contexts can be observable in a wireless network configuration; information that depicts the situation of a user can be entitled context. These context among others include: WI-FI signal availability, signal intensity, network traffic, geography of WI-FI users, surrounding buildings, structures and vegetations, wireless access points, coverage areas, devices used such as PDAs, mobile phones, laptops, desktops, tablets etc, others include the amount of data transferred, capacity, performance etc. These and many more are set of measurable characteristics used in defining the state of a given entity. These entities may either be a network element or an end user terminal. Some of these entities are equipped with context sensing instrumentation which allows them to record and store specific values about the characteristics that they are designed to monitor.

Once this context information is collected by sensing entities, it is transferred to those which are co-located with context manager.

In essence, as far as network side is concerned, entities such as Base Stations (BS), Access Points (APs) and Cache Controllers (CCs) are always accounted for.

Typically variables for defining the context of a wireless network are the number of users in coverage, available capacity in the wireless links and wired backbones, transmission powers, available memory and (aggregated) user request and mobility patterns. Furthermore, concerning user terminals, it is the number of networks within communication range and the interfaces they operate on, the cost per bit offered by those networks as well as the type of context requested by the end users and their individual usage and mobility patterns are considered.

Table 1  Conceptual diagram of context in wireless technology

| CONTEXT |
| --- |
| -number of users<br>-available capacity<br>-transmission power<br>-available memory<br>-user request<br>-mobility patterns<br>-number of network in coverage<br>-cost per bit<br>-content requested<br>-etc |

A. CONTEXT MANAGEMENT IN WIRELESS NETWORK

In a wireless environment, managing context becomes another issue altogether. Context information is maintained in order to provide timely information to an adaptation algorithm or manager so that content and the behavior of applications can be altered to suit current conditions. This adaptation may be transparent or visible to applications that run on top of this adaptation layer (the goal is to provide a completely transparent mobile architecture).

In a wireless environment, the attributes that constitute context information are always changing. For instance, a user might move from a high bandwidth network to a low bandwidth network; more memory might be slotted into the device; any number of attributes that make up the policy of an organization could change.

In order to manage context information, there needs to be a way we can describe and store it, and there needs to be a way to update this information and also justifies the choices made by referring to the significant constraints imposed by a wireless environment. A further aim of context management is to enable inference of contexts from other contexts. This feature is really what sets apart the context manager from other systems developed to this point.

Adaptation Manager of a wireless architecture uses the information provided by the Context Manager to make decisions about how to modify content or change some aspect of an application's behavior.

In a wireless architecture, there are many sources of information that must be gathered and presented in a usable form to the other components. Information can come from hardware, software, users, location managers, the organization (the enterprise or institution), or even from the other components of the wireless architecture. For example, some policy information could be used as context information. The Context Manager's task is to collate all this information and transform it into something that can be used by other components.

There is a Context Manager. Its role is to maintain information pertaining to wireless devices, users, the environment around each wireless device and any other context information deemed relevant. All this information is provided by a set of 'awareness modules'. An example of an awareness module is a location manager that tracks the location of users. Another example is a mobile device itself, which can provide information about its present capabilities. All this context information is collated and made available to the other management components. These requirements lead us to three aspects of context management:

- Context sensing. The way in which context data is obtained.

- Context representation. The way in which context information is stored and transported.

- Context interpretation. The way in which meaning is obtained from the representation [15].

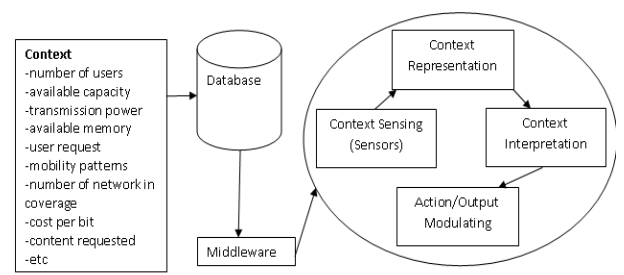

Fig. 2: Conceptual diagram of a Context management process

As represented above this diagram conceptualized the process involved in the collection of context information as entities which is then passed to a storage location interfacing with middleware applications which is in turn passed to the various stages of context processing - sensing, representation of the input context into a form understandable by the context manager, interpretation by the context manager and an eventual execution of the desired action based on the context in consideration.

## IV. CONTEXT MANAGEMENT STRATEGIES IN WIRELESS TECHNOLOGY

Context management strategies are bound to the actual management strategy and refer to collection and distribution of context information in a wireless network, namely how will the context information (monitored in individual node) reach the center of decision making. Context management Strategies in wireless technology is dependent upon the collection of accurate information from the individual nodes. This information (called context information) can be exploited by administrators or automated systems in order to decide on specific management concerns. While traditional approaches for fixed networks are more or less centralized, more complex management strategies have been proposed for wireless networks, such as hierarchical, fully distributed and hybrid ones. The reason for the introduction of new strategies is based on the dynamic and unpredictable nature of wireless networks and their (usually) limited resources, which do not support centralized management solutions.

Various strategies that shall be considered in this research include:
- Centralized strategy
- Hierarchical strategy
- Distributed strategy
- Hybrid (Hierarchical and Distributed) strategy

### A. CENTRALIZED MANAGEMENT STRATEGY

As wireless technology get more ubiquitous, context management administrators may well find they're spending an inordinate amount of time managing them, especially in scenarios with multiple sites. Configuration changes require logging in to the controller at each site, troubleshooting can be a bear, and reporting is all but impossible. That is, unless you have a centralized management strategy.

With centralized wireless context management, the wireless controllers that manage access points in each location all connect to a central management console. Wireless Sensor Networks (WSN) have proved to be useful in applications that involve monitoring of real-time data. There is a wide variety of monitoring applications that can be employed in Wireless Sensor Network. Characteristics of a WSN, such as topology and scale, depend upon the application, for which it is employed. With centralized context management strategy, all of these can be managed from a single control station or server. Businesses often tend to add equipment and software to networks that can make them more complex; the number of systems to manage is sometimes so large that there is a lack of connections between disparate parts. Centralized network management usually makes user access, data storage, and troubleshooting more convenient.

Managing a network generally includes monitoring performance, but security, balancing of processor loads, and traffic management are usually important as well. A server can be centralized to monitor various operational parameters. It can react in response to particular actions or if certain levels of traffic or processing activity are reached. Operational and security policies can also be set in the system so that centralized network management can be performed efficiently. Specialized software for centralized network management often helps to monitor, analyze, and manage all of the components. It can allow an administrator to set policies, assign network equipment for specific resources, and view the performance of individual components or the entire system on a graphical screen. These data can be applied to developing a model of the network, planning for additional capacity, and forecasting how certain variables will affect performance. Network administration not only involves managing data and system operations. One must also be able to control who has access to what. Passwords and user permissions, as well as event logging, are often part of centralized network management too. On some networked systems, it can be possible to load software from remote locations, especially if there is a connection between data storage and a remote server.

There are many software programs that can be used and accessed throughout an enterprise. Centralized context management enables these programs to be installed quicker, and for people to access them. If different parts of the network were disconnected from one another, accessing such systems is often difficult [21].

Centralized context management is typically useful when many applications running in the system interfere with performance. Services that work with voice, video, and data often face such problems, so administrators can adjust network functions to make things more efficient, like with a communications network, for example. By centralizing operations, administrators can better understand regular network functions and also predict how any changes or upgrades will impact the way things work [12].

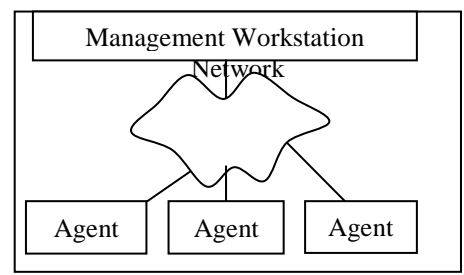

Fig. 3: Centralized context management strategy.

### B. HIERARCHICAL CONTEXT MANAGEMENT STRATEGY

According to Mohsen et al, large enterprise networks span application, organizational and geographical boundaries. In order to cope sufficiently with the unpredictable growth of the number of network devices, structuring networks in logical hierarchies is being employed as a design and deployment principle. Any of the following or a combination of these partitioning criteria can be used: (a) geographical subdivisions; (b) administrative subdivisions; (c) grouping based on different access privileges and security policies; (d) performance-driven network partitioning between multiple management servers. The design of such networks' management systems must consider that the resultant network segments may be better managed as logical, hierarchically structured domains.

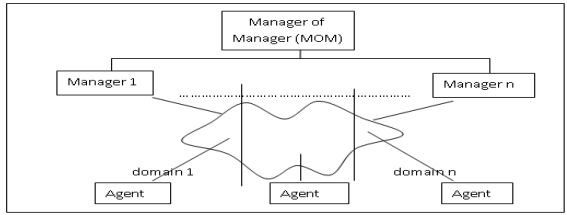

Fig. 4. – Hierarchical context management approach

The hierarchical architecture uses the concept of "Manager of Managers" (MOM) and manager per domain paradigm. Each domain manager is only responsible for the management of its domain, and is unaware of other domains. The manager of managers sits at a higher level and request information from domain managers. In this architecture, there is no direct communication between domain managers. This architecture is quite scalable, and by adding another level of MOM a multiple level hierarchy can be achieved.

Hierarchical network management system also uses the concept of SubManager, similar to the manager-to-manager (M2M) protocol described in Simple Network Management Protocol version 2 (SNMPv2). A SubManager is associated with a few agents, and collects the primitive context information from them, performs some calculations, and produces more meaningful values that can be used by a superior manager. This method significantly reduces the amount of management traffics, because only high-level information is sent to the master manager.

Whenever network operators need more, or high-level, information a downloaded Network Management Procedure (NMP) is dynamically assigned to the system. This allows dynamic reconfiguration at run-time, and removes the need to hard-wire everything at compile time. The NMPs are loaded into submanagers, and are stored in two tables: subMgrEntry and subMgrOps. Each procedure is periodically activated and a "Worker" is created which evaluates the procedure and stores the result into subMgrValue table. This table is read by the manager [10].

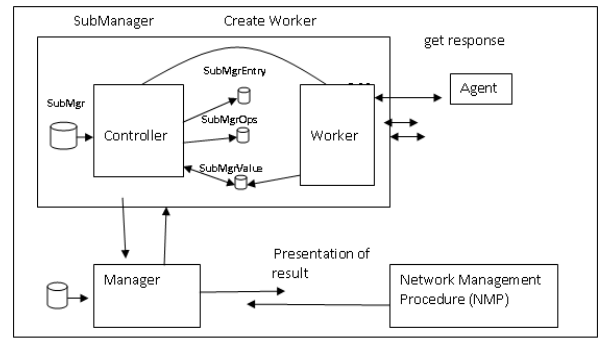

Fig. 5: The Internals of a Context SubManager
Source: Mohsen K, H.W. Peter, 2011

### C. DISTRIBUTED MANAGMENT STRATEGY

Distributed network is a collection of physically separate, possibly heterogeneous computer systems that are networked to provide users with access to the various resources that the system maintains (Abraham S., Peter B. & Greg G., 2006)

From Mohsen et al, a distributed system should use interconnected and independent processing elements to avoid having a single point of failure. There are also several other reasons why a distributed system should be used. Firstly, higher performance/cost ratio can be achieved with distributed systems. Also, they achieve better modularity, greater expansibility and scalibility, and higher availability and reliability.

Distribution of services should be transparent to users, so that they cannot distinguish between a local or remote service. This requires the system be consistent, secure, fault tolerant and have a bounded response time. The form of communication in such systems is referred to as client/server communication. The client/server model is a request-reply communication that can be: synchronous and blocking, in which the client waits until it receives the reply; or asynchronous and non-blocking, in which the client can manage to receive the reply later.

Remote Procedure Call (RPC) is well-understood control mechanism used for calling a remote procedure in a client/server environment. The idea is to extend the use of procedure call in local environment to distributed systems. This results in simple, familiar and general methods that can be implemented efficiently.

The Object Management Group's (OMG) Common Object Request Broker Architecture (CORBA) is also an important standard for distributed object-oriented systems. It is aimed at the management of objects in distributed heterogeneous systems. CORBA addresses two challenges in developing distributed systems: making the design of a distributed system not more difficult than a centralized one; and providing an infrastructure to integrate application components into a distributed system. The distributed approach is a peer-to-peer architecture. Multiple managers, each responsible for a domain, communicate with each other in a peer-system. Whenever information from another domain is required, the corresponding manager is contacted and the information is retrieved. By distributing management over several workstations throughout the network, the network management reliability, robustness and performance increases, while the network management costs in communication and computation decrease. This approach has also been adapted by ISO standards and the Telecommunication Management Network (TMN) architecture. The Management model for ATM networks, adapted by ATM Forum, is based on this approach, too.

Various variant of the distributed strategy has being proposed among whom are: distributed big brother, from the University of Michigan; Distributed Management Environment, from the Open Software [12].

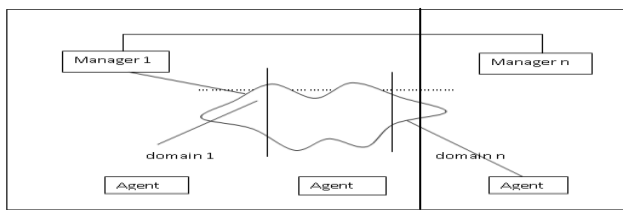

Fig. 6 – A conceptual diagram of distributed strategy

### D. HYBRID CONTEXT MANAGEMENT STRATEGY

Hybrid context management strategy in wireless technology is a paradigm where by context information is being maintained in order to provide timely information to an adaptation algorithm or manager using a combination of hierarchical and distributed strategy to alter the content and behaviour of context information to suit current condition. Hybrid context strategy tends to combine hierarchical and distributed management strategy to achieve thorough and all encompassing processes of collection, exchange and processing of context information. A combination of hierarchical and distributed context management strategy in wireless network is also known as the Network strategy or architecture.

This architecture uses both manager-per-domain and manager of managers concepts, but instead of a purely peer-system or hierarchical structure, the managers are organised in a network scheme. This approach preserves the scalibility of both systems and provides better functionality in diverse environments [12].

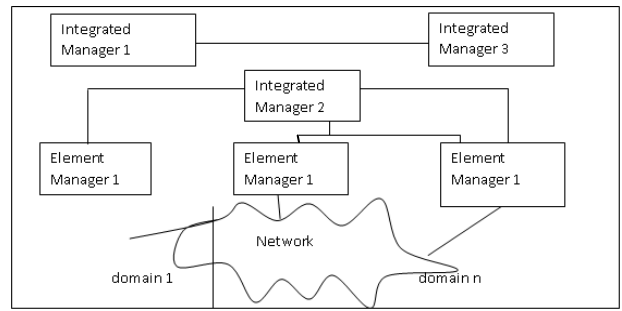

Fig. 7: Hybrid context management strategy Architecture

### V. CHOICE OF STRATEGY

The choice of which strategy to be used in a particular situation depends on the scenario in which the context information will be managed, it is expedient that in the choice of strategy, an analysis of the scenario should be done critically and decision of choice being made in terms of performance and efficiency. The goal is to examine which strategy is more efficient for what scenario e,g. for high mobility scenario, which solution might perform better.

There are a number of metrics that can be used to determine if a network management application is more appropriate, a centralized or distributed system. This metric is shown in the figure below.

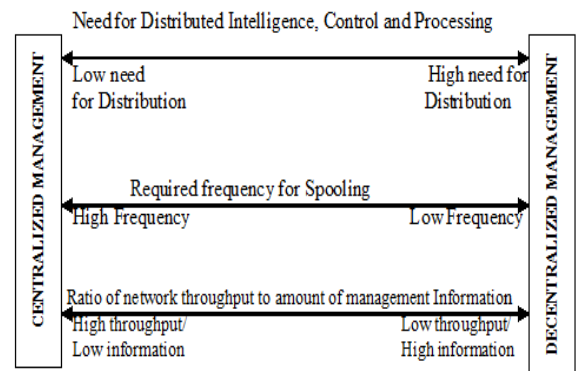

Fig. 8: Metrics for the choice of strategy
Source: Mohsen Kahani, H.W. Peter Beadle

### VI. COMPARISM OF THE VARIOUS CONTEXT MANAGEMENT STRATEGIES

In this section a comparison is made to show the advantages and drawbacks of the discussed systems. We focus our comparison on the performance of these systems for several requirements of distributed network management systems, such as polling method, communication between layers of management, extensibility and flexibility.

Table 2 Comparism of the various strategies

| | Centralized Management | Distributed Big Brother | Distributed Management Environment | Hierarchical Network Management | Management by Delegation |
|---|---|---|---|---|---|
| Architecture | Centralized | Hierarchical | Hierarchical | Hierarchical | Distributed/ Hierarchical |
| Communication method | N/A | Contracting Protocol | RPC | Client/Server | Delegation Protocol |
| Polling method | Direct | Indirect (via group managers) | Indirect (via object servers) | Indirect (via group Submanagers) | Indirect (via elastict servers) |
| Polling Interval | High | Low | Low | Low | Low |
| Autonomy | N/A | Good | Fair | Fair | Very good |
| Extensibility | Low | Medium | High | High | Very High |
| Flexibility | Low | High | Medium | Medium | Very high |

Source: Mohsen Kahani, H.W. Peter Beadle

## VII   CONTRIBUTION OF WORK

This work has being able to analyze some strategies that can be employed by context application developers to effectively manage context information in wireless network, this has being achieved by making a detailed comparison of these strategies in order to make informed decisions in the right choice of strategy to be used in a specific environment and in a combination of various environment.

## VIII   FURTHER WORK

Having outline the theoretical framework for effective management of context information in a wireless network, future work shall be geared towards developing a context reminder that reminds students in a geographical area of the lectures they are suppose to attend on a specific day and after a specific time interval a call is automatically place to the course lecturer reminding him/her of the schedule class and also retrieve his current location using G.P.S technology and propagating the information back to the students
.
Efforts shall also be made using any of the most optimal strategy to develop context aware application that can detect the presence of human beings approaching crude oil pipelines in Nigeria, this will go a long way to help in detecting pipeline vandals in my country and secure our pipelines.

Lastly a context aware application that will use the Hybrid strategy to detect illegal arms and amunitions including illegal explosives shall also be thought of.

## IX. CONCLUSION

This work has been able to bridge the gap in knowledge of context management strategy as regards wireless network, it has uncovered the strategies to be employed in managing context information that comes into a context aware system, it has also been able to make detail a comparison of the various strategy and has given suggestions on the choice of strategy for optimal and efficient context management, of particular interest is the Hybrid strategy that can adapt to various environment owing to the fact that context applications are mostly environment specific.